\documentstyle[preprint,prl,oldlfont,aps]{revtex}
\begin{document}
\draft
\title{Chain length scaling of protein folding time.} 
\author{A. M. Gutin, V. I. Abkevich, and E. I. Shakhnovich}
\address{Harvard University, Department of Chemistry\\
12 Oxford Street, Cambridge MA 02138}
\maketitle
\begin{abstract}
Folding of protein-like heteropolymers into unique 3D structures is investigated using Monte Carlo simulations on a cubic lattice. We found  that folding time of chains of length $N$ scales as $N^\lambda$ at temperature of fastest folding.  For chains with random sequences of 
monomers $\lambda \approx 6$, and for 
chains with sequences designed to 
provide a pronounced minimum of energy to their ground state conformation 
$\lambda \approx 4$. Folding at low temperatures exhibits an Arrhenius-like behavior with the energy barrier $E_b \approx \phi |E_n|$, where $E_n$ is the energy of the native conformation. $\phi \approx 0.18$ both for random and designed sequences.
\end{abstract}
\narrowtext
The number of all possible conformation (shapes) 
of a protein of $N$ amino acids scales as $z^N$ where $z$ is the number of conformations per one amino acid residue. Thus, random search in the conformational space for the native, biologically active 
shape would require folding time $t \sim z^N$ which is unrealistic 
for a typical protein with $N \sim 100$ (Levinthal paradox).
  In fact, real dynamics of a protein is far 
from  random search. The native structure is 
believed to represent a pronounced 
minimum of energy \cite{Go,GSW,SG93,sali}, 
so that folding process is driven by 
decrease in  energy (opposed by entropy decrease). Moreover, 
it is clear that folding time should 
strongly depend on folding conditions. 
At very high temperature all 
conformations are equally favorable 
and  folding time should indeed be close 
to Levinthal's estimate $z^N$. At very 
low temperature one should expect 
Arrhenius-like slowing-down caused by 
energy barrier(s) on a folding 
pathway. Thus, there should be some 
finite temperature optimal for folding 
in the sense that folding  
is fastest at this temperature. 
Such temperature dependence of the 
folding time was observed indeed 
in a number of computer simulations for different models \cite{dill92,AGS94,dill94,BOSW}.

Much less is known about length dependence of the folding time. It is clear that folding time should grow when chain length increases
but the actual dependence is not known. Though there are some attempts to estimate the folding time from scaling and other simple arguments \cite{SG89,BW89,thirumalai}, these estimates 
are based on phenomenological, rather than microscopic, considerations. Experimental data are not readily available to elucidate this 
important point. The problem here is that  folding can involve some 
kinetic events which are specific to a given protein, 
such as isomerisation of prolines and formation 
of disulphide bonds. In addition, large proteins 
typically consist of few domains with independent, 
to some extent, folding.

In the present paper we study the length dependence 
of folding rate using Monte Carlo simulation of folding 
on a cubic lattice. Despite its simplicity and computational efficiency,
this model proved useful in elucidating such crucial,
experimentally observed, 
features of protein folding as cooperativity
\cite{PRL94,HS,JMB95}, nucleation mechanism \cite{NUCLEUS,F95,CONSERV}
and parallel pathways \cite{DOBSON,sali,AGS94,BOSW}

Both chains with random 
sequences of amino acids and chains with 
sequences designed to provide a pronounced 
energy minimum for a particular 
compact conformation \cite{SG93} are analyzed and compared.

Protein chain is modeled as a self-avoiding 
walk on an infinite cubic lattice \cite{AGS94}. 
Energy of a given conformation of a chain is given by
\begin{equation}
H= \sum_{i<j} U(a_i, a_j) \Delta_{ij}
\end{equation} 
where $\Delta_{ij}=1$ if monomers $i$ and $j$ are in contact and $\Delta_{ij}=0$ otherwise. Monomers $i$ and $j$ are considered to be in contact if they are separated by one lattice bond and $|i-j| \ne 1$. 
Sequence of amino acids is given by 
$a_i$ with $i=1,...,N$, so that 
$a_i$ can take 20 different values 
corresponding to 20 natural amino acids. 
Finally, $U(a, b)$ is the energy of a 
contact between amino acids $a$ and $b$. 
In this work we use interaction matrix 
$U(a, b)$ obtained in Ref. \cite{MJ}. 
from statistical analysis of protein 
structures. A number of previous studies suggest
\cite{NUCLEUS,CONSERV} that the results 
do not depend on a particular choice of 
interaction matrix. Dynamics of a chain 
is modeled by a standard Monte Carlo 
procedure for a cubic lattice \cite{MC}.

First, we studied folding of chains 
with random sequences of monomers. 
5 random sequences 
were generated for each 
chain length in the range from 10 to 50. 
A long Monte Carlo simulation was performed 
for each of these sequences,  
to identify the conformation with the 
lowest energy. Fig.1 shows the lowest 
\marginpar{Fig.1}
energy conformation for one of random 
chains of 40 monomers. 
In order to make 
sure that there are no conformations 
with lower energies, additional 10 
runs starting from different unfolded 
conformations were performed for each sequence. 
In each of these runs a chain 
folded to the putative lowest energy conformation, 
and no conformation with lower energy was encountered.

From 5 random sequences generated for the 
same chain length, we selected one sequence 
which folding rate was the fastest. 
Then for this sequence we studied folding in  
a range of temperatures to determine
the temperature, at which folding rate
was fastest. 
Further, at this temperature, 
we made a more
precise estimate of the mean first passage time (MFPT) to 
the lowest energy conformation by averaging over 50 runs 
starting from different unfolded 
conformations. The MFPT for all 
studied chain lengths is shown 
in Fig.2. It is seen that folding 
\marginpar{Fig.2}
time grows with chain length, and 
the dependence can be well described by a power-law, 
that is $t \sim N^\lambda$ with $\lambda \approx 6$.

Next, we studied 
the length dependence of folding rate
for sequences designed to have their 
native conformations as pronounced 
energy minimum. Such sequences are more likely to
exhibit protein-like behavior \cite{PRL94}. To this end, 
each of the lowest energy conformations 
found previously for random sequences was used 
as a target conformation for  sequence design. 
The design procedure is described in detail 
elsewhere \cite{SG93,JMB95,AGS96}. 
We did not study random sequences longer than 50 units,
while it is feasible to study folding rate 
of designed sequences up to 100 units long \cite{PRL94}. 
To this end we generated random compact conformations to serve as target
structures for sequence design of longer sequences (more than 50 units
in length).
Additional condition
was applied that variation of energies of 
native contacts is low enough, in order to eliminate
multi-domain behavior for longer sequences \cite{AGS96}.
In this way 5 sequences were generated 
for each of studied chain length, 
and the sequence with the fastest 
folding was selected. Length 
dependence of the folding time at 
optimal temperature for the designed 
sequences is presented in Fig.2 for  
chain lengths ranging from 10 to 100. It can be 
seen 
that for short chains folding time 
of  designed sequences is about the 
same as that of random sequences. For 
longer chains, though,  folding of  designed 
sequences is much faster than folding 
of random sequences. Overall, length 
dependence of the folding time for designed sequences 
is well 
described by a power law $t \sim N^\lambda$ 
with $\lambda \approx 4$.

For comparison, we also studied the 
model introduced by Go and coworkers \cite{Go}. In this model
energy of a given conformation is given by 
\begin{equation}
H= -\frac{1}{2} \sum_{i<j} \Delta^N_{ij} \Delta_{ij}
\end{equation} 
where $\Delta^N_{ij}=1$ if monomers $i$ and $j$ are in contact in the native conformation and $\Delta^N_{ij}=0$ otherwise. In other words, in the Go model all the native interactions are favorable and all non-native interactions do not contribute to the energy. In some sense, the Go model corresponds to ideally designed sequences.

For the Go model the same conformations as for designed 
sequences were used as native. 
Length dependence of the folding 
time at optimal temperature for 
the Go model is presented in Fig.2 
for chain lengths from 10 to 175. 
Again, the dependence is fitted very 
well by a power law 
$t \sim N^\lambda$ with $\lambda \approx 2.7$.

Fig.3 shows the inverse temperature
\marginpar{Fig.3} 
dependence of the folding time for  
the sequence designed to fold to the 
native 
conformation shown in Fig.1. It 
has a clear minimum at 
some optimal temperature. 
All the scaling dependencies for the folding rate
discussed above were obtained at the conditions
corresponding, for each sequence, to such an optimal temperature.
What happens at lower temperatures?
It is clear from Fig.3 that
at low temperatures logarithm of  
folding time depends linearly 
on inverse temperature 
exhibiting an Arrhenius-like 
behavior $t \sim \exp(E_b/T)$. 
This suggests that at low temperature
folding is an activated process which entails overcoming
of an 
energetic barrier $E_b$.
Similar dependencies were obtained 
for all  studied chains and 
the corresponding energy barriers 
$E_b$ were determined for each sequence from the 
slopes of
Arrhenius-like branches of these dependencies.  
The result is 
presented in Fig.4 where $E_b$ is
\marginpar{Fig.4} 
plotted 
as a function of energy of 
the native conformation, $E_n$. 
It can be seen that energy of 
the low-temperature barrier scales 
linearly with energy of 
the native conformation with  
coefficient $\phi  \approx 0.18$. 
Surprisingly, all the data for 
random and designed sequences 
and for the Go model fall on the 
same line suggesting
a universal scaling behavior for the low-temperature
energy barrier. This observation can be explained as follows.

The main reason that folding is slow at low temperatures 
is that the chain gets stuck in some 
low energy misfolded conformations which needs to
be unfolded before chain proceeds further. 
In the case of random sequences 
such trapped misfolded conformations are 
typically very different in 
structure from the native conformation 
\cite{SG89a}. Nevertheless, the energy 
of the misfolded traps is 
expected to be close to the energy 
of the native conformation $E_n$ \cite{GAS95}. 
In the case of designed 
sequences and Go model
the native state has much lower energy 
than any conformation structurally unrelated 
to the native state. 
Therefore, for designed sequences, 
 low-energy conformations 
serving as kinetic traps, 
have considerable structural similarity 
with the native state.
 As a result, 
the energy of  deepest kinetic traps 
are again close to $E_n$. In order to 
escape from such a misfolded state a 
chain has to break some part of the 
misfolded conformation. Our simulations 
suggest that both random and designed 
chains on a 
cubic lattice have to rearrange a fraction $\phi$ of the 
whole structure (independently of its size), in order to escape
from a trap. We should note though that the actual estimate of
the number of bonds to be broken to escape from the native
state may depend on lattice and dynamic algorithm used (e.g. move set).
We cannot exclude that this barrier will be dramatically
diminished in off-lattice model, so that  low-temperature
behavior in off-lattice model may be significantly different.

As we pointed out above, the direct comparison 
between our main 
result and experiments
on real proteins is a challenging task for experimentalists.  
This will require
careful choice of proteins to study as model systems and
conditions at which their folding is fastest. We hope that the 
presented
results will stimulate such experiments.

The major finding of this work is that the chain length dependence
of folding rate is relatively weak (much weaker than exponential, suggested
by Levinthal argument and some phenomenological estimates \cite{BOSW}).
This fact is of fundamental importance since it explains, from the
folding perspective,
the wide range of lengths of existing proteins
(roughly 50-1000 aminoacids). Exponential dependence of folding
time on protein length would make folding of longer chains prohibitively
slow. Another important application of the present result
is that it can provide crucial and delicate test of
the existing and future 
kinetic theories. 
Phenomenological analysis based on the Random Energy Model (REM)
predicts exponential dependence of folding rate on system size
at all temperatures \cite{BOSW}.  
This is in contrast with simulation results and probably 
points out
to inapplicability of the REM
(and perhaps mean-field models in general) to tackle kinetic issues.
This can be understood from the general perspective
that folding transition is cooperative (first-order like).
Such transitions follow nucleation mechanism at which
the transition state is highly inhomogeneous,
consisting of islands of the ``new'' phase in the
sea of ``old'' phase \cite{LP}. Such
an inhomogeneous 
distribution (representing 
the least-activation path) 
explains the weak size dependence of the
rate of first-order transitions. It 
cannot be described by any global homogeneous
order parameter.
REM-based phenomenological models \cite{BOSW}, or 
approaches, based on Kramers theory \cite{SOW},  use a global order parameter 
(the number of native-like contacts, Q) as kinetic reaction coordinate.
Such approaches 
miss the 
inhomogeneity of the transition state
 and thus  predict 
 exponential size dependence of protein folding rate.

A possible physical explanation of the power law rate dependence, 
for designed sequences is close to
the arguments presented recently by Thirumalai \cite{thirumalai}.   
Since folding  is cooperative (first-order like) process 
\cite{PRL94,HS}, the 
transition
state is reached when nucleus of the
folded conformation is formed 
\cite{NUCLEUS,F95,CONSERV}. 
The  power-law length scaling of the folding 
rate  implies that nucleus does not grow with chain length.
In this case the length dependence of the folding rate
appears due to entropic cost of loop
closure around the folding nucleus. 
The free energy of loop closure depends logarithmically on its length
\cite{degennes}, and this factor, combined with power-law
length dependence of polymer relaxation time \cite{degennes} apparently translates into the overall
power-law dependence of the folding rate. 

The difference between
random and designed sequences is crucial .  First of all we see
that designed sequences fold faster at their respective 
optimal folding temperatures and the difference
in folding rates between random and designed sequences
becomes more pronounced as chains become longer
(due to different exponents in the scaling laws
describing length dependencies).

Real proteins must not only reach their native state
fast but also be stable in their 
native conformations. This points out to another crucial difference
between random and designed sequences: designed
sequences of different lengths are stable in the native state
at the
temperature of their fastest folding
while longer random sequences get less stable at their corresponding
fastest folding temperature. (Data not shown).

Finally, we note that 
while power law scaling fits 
our data best, the range of chain lengths,
which we studied, spanned only about an order of magnitude. 
This length range is most limited 
for random sequences; for them
we cannot rule
out a weak stretched exponential, rather than power
length scaling \cite{thirumalai}. 
We believe that microscopic analytical theory
of protein folding dynamics 
can give satisfactory answer to the important question
of length dependence of protein folding time.
It is our hope that presented results will stimulate
the development of such theory.

\acknowledgements This work was supported by the David and Lucille Packard Foundation and NIH. Interesting discussions with 
D.Thirumalai and A.Grosberg are gratefully acknowledged.

\pagebreak

\pagebreak

{\bf   Figure  Captions}

{\bf Fig.1} Lowest energy conformation for a random sequence of 40 amino acid residues.

{\bf Fig.2} Dependence of the folding time $t$ on the chain length $N$ for random sequences (black circles), for designed sequences (gray circles), and for the Go model (white circles). Folding time was estimated by averaging over 50 folding runs for each of the sequences.

{\bf Fig.3} Dependence of the folding time $t$ on the temperature $T$ for a random sequence of 40 residues with the ground state shown in Fig.1. Straight line approximates low-temperature part of the dependence. The slope of the line is the energy barrier $E_b$.

{\bf Fig.4} Dependence of the energy barrier $E_b$ on the energy of the native conformation $E_n$ for random sequences (black circles), for designed sequences (gray circles), and for the Go model (white circles). The line shows the best fit by $E_b=\phi |E_n|$ with $\phi=0.18$.

\end{document}